\documentstyle[12pt,aps,epsfig,tabularx]{revtex}
\begin{document}            
\title{Formation of magnetic characteristics \\
and hyperfine fields in metal-metalloid alloys}
\author{A. K. Arzhnikov and L. V. Dobysheva}
\address{Physical-Technical Institute, Ural Branch of Russian Academy 
of Sciences, Izhevsk, Russia}
\maketitle

\begin{abstract}
This work deals with the analysis of peculiarities of formation of the
hyperfine fields (HFF) at the Fe nuclei in disordered alloys metal- 
metalloid using the "first-principles"  calculations. Some 
phenomenological models and justification of their usage for the 
interpretation of the experimental HFF distributions are discussed.
\end{abstract}



\vskip .5cm
1. Introduction

The development of technologies for preparing materials for modern 
electronics requires a detailed knowledge of different local physical 
quantities at the atomic level. In the experimental physics the 
possibility to measure the local characteristics at such a scale are 
very limited. Among few existing methods, of special accuracy are the 
nuclear ones (for example, the Mossbauer spectroscopy) that measure  the
hyperfine magnetic fields (HFF) at nuclei. It is widely believed that
these techniques can reflect some chemical and topological features of
the atomic surroundings of the excited nuclei.  Indeed, in numerous
studies the spectra are interpreted in terms of phenomenological
models considering only the nearest atomic environment (see e.g. [1-3]).
There is no doubt that such a description is often useful and efficient.
We think, however, that there are some cases when such a simplified
approach is not justified. Moreover, for the spectra of transition
metals and alloys on their basis, the successful description within the
framework of such models is rather an exception than a rule, and every
time additional arguments are needed to justify these limitations. 

In this work we analyze the peculiarities of formation of HFF in 
disordered alloys of Fe and metalloid using the "first-principles" \
calculations. One may consider an HFF as a sum of three main terms:
$H^{core}$,  $H^{val}$ and $H^{orb}$ which are the core, valence and
orbital  contributions, respectively. This standard division is based on
the  physical features that determine the magnitude of these 
contributions, $H^{core}$ and $H^{orb}$ being mostly dependent on the 
topological and chemical peculiarities of the nearest neighboring only, 
whereas $H^{val}$ is significantly influenced by the atoms of the 
second and more distant spheres. At the same time, in disordered alloys 
the effect of the distant spheres on $H^{val}$ may be considerably
suppressed due to  the great localization of the valence electrons.
Owing to this, in some  cases [1-3] the HFF distributions can be
successfully described by  phenomenological models of the
Jaccarino-Walker type where the  magnitudes of the local magnetic
moments (LMM) and HFF are taken to be  dependent on the number of
metalloid atoms in the nearest neighboring.  Often such descriptions are
surprisingly accurate and efficient. One  must bear in mind, however,
that the phenomenological models have  fitting parameters, and there are
various examples in the history of  sciences when the accuracy and
efficiency were illusory, being only a  result of manipulation with the
number and magnitudes of the parameters. 

To avoid such a trap, one must realize how to remove the contradiction 
between the localized character of the description of HFF and the 
itinerant behavior of the d- and sp- electrons in metals. At present, 
the solid state theory based on the "first-principles" calculations 
allows one not only to justify the use of some phenomenological models, 
but also to make them more reliable by defining the parameters from the 
"first-principles" calculations. 

In this paper, we analyze the behavior of $H^{core}$, $H^{val}$ and 
$H^{orb}$ using the "first-principles" \  calculations performed for the 
ordered alloys $Fe_{100-c}Me_c$ ($Me=Si, Sn$), namely: $Fe_{31}Si$,
$Fe_{15}Si$, $Fe_{31}Sn$, $Fe_{15}Sn$, $Fe_{26}Si_{6}$, $Fe_{25}Si_{7}$,
$Fe_{24}Si_{8}$, $Fe_{23}Si_{9}$ and $Fe_{22}Si_{10}$,  and experimental
data [4].  The calculations were carried out by the full-potential
linearized augmented plane wave method (FLAPW) using the WIEN-97 program
package [5]. Some results are presented in Table 1.

The systems were simulated on a bcc lattice which in the disordered
alloys under consideration is retained within a wide concentration range
[1]. The lattice parameters were chosen in accordance with the
experimental values of the alloys with corresponding concentrations 
($c=3.125, 6.25, 18.75, 21.875, 25, 28.125$ and $31.25 at.\%$). It
should be mentioned that even at small concentrations the bcc lattice is
somewhat distorted due to the repulsion/attraction by the Sn/Si atom 
magnetic of the surrounding Fe atoms. As shown in our paper [6], the
changes in characteristics because of this relaxation are insignificant.
Though the results in Table 1 were obtained with allowance for this
relaxation, we do not discuss it here.

\vskip .5cm
2. The local magnetic moment

Hereinafter, the local magnetic moment (LMM) at Fe atom $i$ means the
spin magnetic moment over the muffin-tin (MT) sphere, $M_i^d$. The
average  magnetic moment over the unit cell is  $M=\sum_i
(M_i^d+M_i^{orb}) + M^{int}$, where $M_i^{orb}$ is the orbital magnetic 
moment, $M^{int}$ is the spin magnetic moment outside the MT spheres. 
The main contribution to the LMM comes from the d-electrons that are
almost entirely inside the MT-sphere, whereas $M_{int}$ is formed by the
s- and p- electrons and has a small negative value as compared to the
LMM. We neglect $M^{orb}$, as its variations with concentration and the 
metalloid atoms arrangement are small and amount to less than 1 \%  [7].
The calculated values are two times less than the experimental ones. We
believe that the actual changes of $M^{orb}$ do not exceed twice the
calculated ones. It should be noted that the neglect of $H^{orb}$
requires  some caution: in what follows we shall show that the changes
in  $H^{orb}$ may be quite significant amounting to 10 $\div$ 15 \% of
the  average HFF. Mention the principal features of the LMM formation.

A. The main peculiarities of the magnetic moment depend weakly on the
metalloid type and, as shown in [8], is governed by the lattice
parameter, a. At equal concentrations, the lattice parameter of the Sn
alloy is greater than that of the Si alloy, so the magnetic moment in
the first case is greater.

B. The LMM of the Fe atom closest to the impurity is of the smallest among
other values. As noted in [8], this difference is defined by
the competition between two mechanisms: the LMM reduction due to flattening
of the d-band because of the s-d hybridization that is the strongest near
the impurity, and the LMM increase due to narrowing of the d-band because of
a decrease of the wave function overlap. There also exists third mechanism
of the LMM reduction due to the difference in the impurity-potential
screening by d-electrons (the difference in pushing out the impurity levels
by the bands with spin up and down) [7].

C. The LMMs are concentration dependent. So the LMM of the Fe atom
closest to the Si atom is 2.181 $\mu_B$ at c=3.125 at.\% ($Fe_{31}Si$), and
2.262  $\mu_B$ at c=6.25 at.\% ($Fe_{15}Si$) (Table 1). 

At the same time, our "first-principles" calculations and the
mathematical treatment [9-10] testify that, to an accuracy of $10 \div
15 \%$, the LMMs in disordered  alloys are independent of concentration
and governed by the number of  nearest metalloid atoms only. 

D. It is natural to expect that, with a large number of nearest metalloid 
atoms the LMM will considerably decrease down to zero, due strong s-d 
hybridization [10]. 

\vskip .5cm
3. Hyperfine magnetic fields at nuclei

The program package WIEN-97 allows one to calculate the interaction
between the nucleus magnetic moment and the spin and orbital magnetic
moments of the electron subsystem. The spin dipole contributions are
small (~2 $\div$ 3 kGs), they are suppressed due to the symmetry
relations, and the main contribution comes from the spin polarization at
the nucleus (Fermi-contact interaction), $H^{core}+H^{val}$, and the
orbital magnetic moment. For the core-electron polarization the simple
relation $H^{core}_i =\gamma^s Md_i$ is satisfied, where $\gamma^s$ does
not depend to a high accuracy on neither the metalloid type nor its 
concentration, and is determined only by the approximation for the
exchange-correlation potential.  Here we use the GGA approximation [11]
which gives $\gamma^s\approx 123  kGs/\mu_B$. Due to the simple relation
between the LMM and $H^{core}$  and owing to the possibility of the
rough description of the LMM as a function of neighboring metalloids 
number discussed in Section 2 point C, if
$H^{val}$ is not too large one can use the Jaccarino-Walker-type
phenomenological  models while interpreting the HFF distributions of
disordered metal-  metalloid alloys, but the accuracy in this case
cannot be high. 

The proportionality $H^{orb}_i =\gamma^{orb} M_i^{orb}$ is fulfilled in
a somewhat worse way, but still rather satisfactorily. Note, however,
that $\gamma^{orb}$ is positive and about five times larger in magnitude
than $\gamma^s$. Hence, if the changes of $M_i^{orb}$ affect the
magnetic moment only slightly, the changes in $H_i^{orb}$ with 
allowance for the actual magnitude of $M^{orb}$ may amount to 20 kGs
even at low concentrations (see Table 1). 

$H_i^{val}$ behaves in a more complicated way. This is primarily
associated with strong delocalization of the s- and p-like electrons
that interfere at sites with different magnetic and charge properties,
and therefore the $H_i^{val}$ behavior cannot be in fact quantitatively
predicted in disordered systems. However, we succeeded in revealing some
qualitative regularities supported by experimental evidence. We analyzed
this  contribution using the simple functional relation 
$$
H_i^{val}=F_0 M_i+ \sum_j F_j N_j,\eqno(1)
$$ 
where $F_0$ is a constant describing the 
polarization of the valence electrons of the Fe  atom at site $i$ by its
own magnetic moment, $F_j$ are factors  describing the change of
polarization at atom $i$ by the metalloid atom in sphere $j$,  $\sum_j$
is summation over all coordination spheres ($j>0$), $N_j$ is the number of 
metalloid  atoms in sphere $j$. Fig.1 shows the values of $F_j$ for  the
calculations of different sets of  alloys. The first curve was obtained 
by solving the equation system Eqn. (1) by the least-squares method with 
calculated values $H_i^{val}$ and $M_j$ for the alloys  $Fe_{26}Si_{6}$,
$Fe_{25}Si_{7}$, $Fe_{24}Si_{8}$, $Fe_{23}Si_{9}$ and $Fe_{22}Si_{10}$, 
the second curve was calculated for the data for the alloys $Fe_{31}Si$
and  $Fe_{15}Si$, the third curve was received for $Fe_{31}Sn$ and
$Fe_{15}Sn$ alloys. One can see that $F_j$ is of oscillating  nature and
depends on the metalloid type and concentration. This  dependence reveals 
itself mostly at close distances, which leads to an  amazing
conclusion: $H^{val}$ may be considered, to some extent, as a linear 
superposition of the changes in spin polarization by distant atoms,
whereas the polarization by close atoms can hardly be regarded  as
linear because of a significant re-distribution of the charge  density
which is not linear.

\begin{figure}[t]
\epsfxsize=15cm     
\centerline{\epsfbox{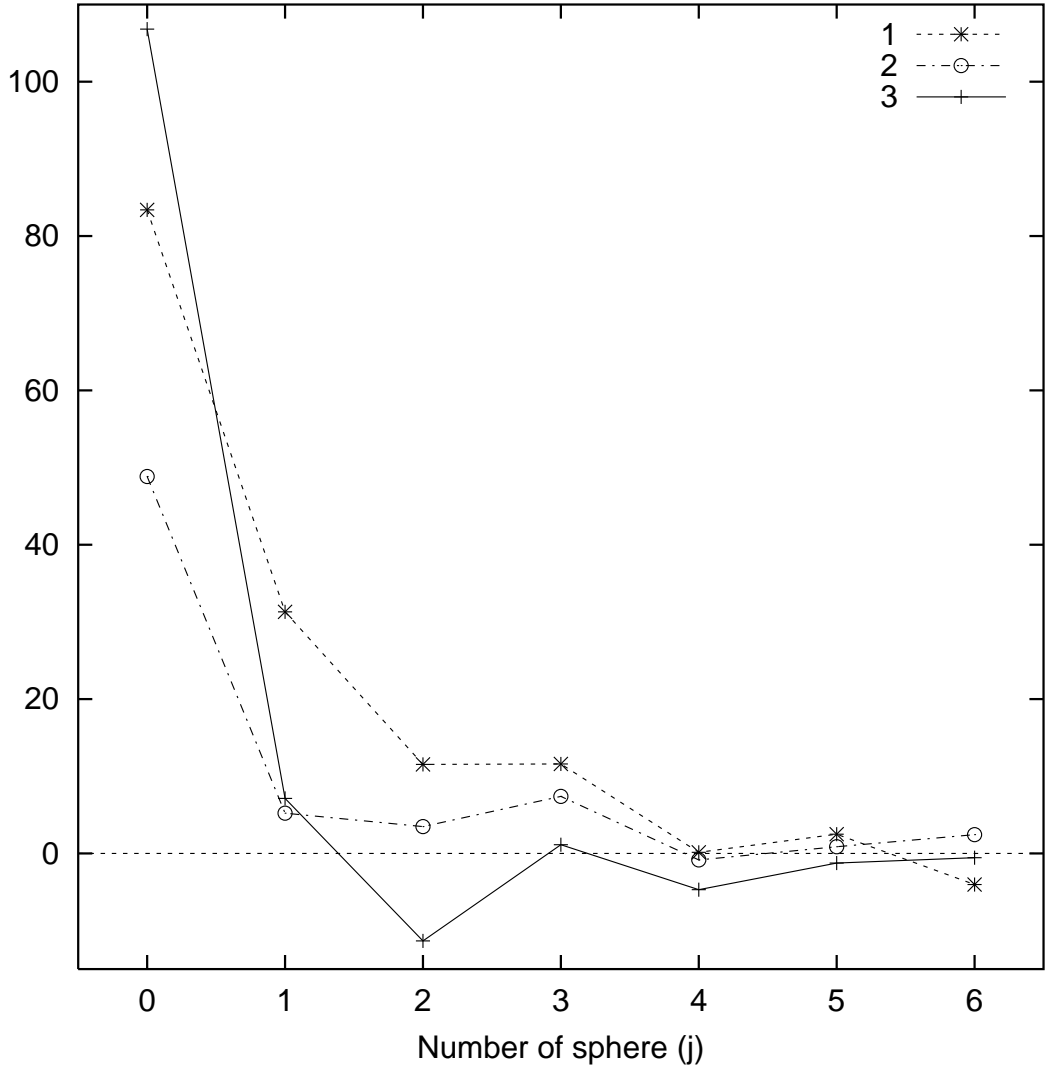}}  
\caption{The parameters of the model (1), $F_j$
($kGs$), as a function of sphere number}

\end{figure}

Thus, it is clear why the use of the simplest RKKY (Ruderman- Kasuya-  
Yosida- Kittel) interaction for the description of HFF resulted
sometimes  in reasonable conclusions for superlattice. First, for the
realistic sp-electron  densities, this interaction has an oscillation
period close to our  result ($2 \div 3$ interatomic distances, Fig.1).
Second,  the spin polarization by the distant metalloid atoms may be
considered  additive. In the superlattices (e.g. in the $DO_3$ type
ordered alloy $Fe_{75\pm c}Si_{25\mp c}$), the possibility of a
configuration of impurity (here, by an impurity is meant an atom of $Fe$ 
or $Si$ which appears at a "wrong" \  site as compared to the totally 
ordered $Fe_3Si$ alloy) atoms arrangement around the $Fe$ atom with
large number of  the impurity atoms in the second and third spheres may
be high even at  low concentration of the impurity atoms. In the light
of the above, the models where the spin polarization is  considered as
an additive contribution from the RKKY polarization by  the first
sphere, are doubtful. 

So far, our statements were mainly  based on the "first-principles"
calculations of the ordered alloys,  while the disordered one were
discussed. It should be mentioned that  the problem of quantitative
calculations of HFFs in disordered alloys  remains unsolved, that is why
we can examine the effect of disorder only qualitatively, using the
model systems.

In our paper [8], the effect of disorder was investigated by 
averaging the RKKY interaction whose parameters were determined from the
"first-principles" calculations. Now we realize that such an  approach
should be used with caution. However, the result of this  paper [8]
that the decrease of the HFF magnitude should be larger  than the
decrease in the magnetic moment with disordering (i.e. with  the
increase in metalloid concentration), holds true if the behavior of  the
polarization by the third and most distant spheres being averaged in 
disordered alloys, is similar to that of Fig.1 (the positive
polarization by the third sphere). 

Note that the decrease in $H_i / M_i$ is also determined by the 
increase in the positive contribution of $H^{orb}$ with the increase of
the disorder degree [12]. It is difficult to separate these 
contributions and to answer unambiguously why the change of this ratio
is  experimentally observed.

\vskip .5cm
Acknowledgements

One of the authors (A.K.A.) is thankful to the Organizing Committee of 
the conference EMRS-2001 and Prof. H.Dreysse for the financial support. 
The present work was supported by Russian Foundation for Fundamental
Researches, Grant No 00-02-17355.

\vskip .5cm
References

[1] E.P. Yelsukov, Phys. Met. Metallogr. 76 (1993) 451. 

[2] N.N. Delyagin et al, Journal Eksp. i Teoret. Phys. 116 (1999) N 1(7),
130. 

[3] B. Rodmacq et al, Phys. Rev. B 21 (1980) 1911. 

[4] E.P. Yelsukov, Phys. Met. Metallogr. 85 (1998) 307.

[5] P. Blaha, K. Schwarz, J. Luitz. WIEN97, A Full Potential Linearized
Augmented Plane Wave Package for Calculating Crystal Properties (Karlheinz
Schwarz, Techn. Universit"at Wien, Austria). (1999) ISBN 3- 9501031-0-4.

[6] A.K. Arzhnikov, L.V. Dobysheva, Phys. Rev. B 62 (2000) 5324.

[7] A.K. Arzhnikov, L.V. Dobysheva, ArXiv: cond-mat /0101033 (2000).

[8] A.K. Arzhnikov, L.V. Dobysheva, F. Brouers, Phys. Sol. St. 42
(2000) 89.

[9] A.K. Arzhnikov, L.V. Dobysheva, E.P. Elsukov, A.V. Zagainov, JETP 83
(1996) 623.

[10] A.K. Arzhnikov, L.V. Dobysheva, JMMM 117 (1992) 87. 

[11] J.P.Perdew et al, Phys. Rev. Lett. 77 (1996) 3865.

[12] A.K. Arzhnikov, L.V. Dobysheva, Izvestiya Akademii Nauk. 
Ser.Fiz. 65 (2001) 985. (in Russian).

\newpage
\begin{table}
\caption{The results of the calculations. Configuration of impurities 
in the Fe-atom environment, [nm...], denotes the number of metalloid atoms 
in the first (n), second (m) etc. spheres. $N_{Fe}$ is the number of such
Fe atoms in the unit cell. }
\vskip .4cm
\begin{tabular}{ccccccc}
\hline
   &$[nm...]$, $N_{Fe}$ & $M^d$, $\mu_B$ & $H^{core}$,kGs 
& $H^{val}$,kGs & $M^{orb}$, $\mu_B$ & $H^{orb}$,kGs \\
\hline
$Fe_{31}Sn$, & [1000], 8& 2.246 & -277.50 & -37.59 & .0489 & 27.94\\
a=21.804a.u. & [0003], 8& 2.507 & -310.35 & -30.94 & .0483 & 26.41\\
             & [0100], 6& 2.432 & -300.17 & -34.84 & .0482 & 26.62\\
             & [0020], 6& 2.425 & -299.79 & -22.39 & .0479 & 26.28\\
             & [0000], 2& 2.391 & -295.13 & -34.34 & .0441 & 23.66\\
             & [0000], 1& 2.423 & -299.68 & -24.17 & .0470 & 25.15\\
\hline
$Fe_{15}Sn$, & [1003], 8& 2.442 & -301.01 & -20.25 & .0548 & 29.77\\
2a=21.985a.u.& [0200], 3& 2.530 & -312.64 & -29.97 & .0535 & 29.08\\
             & [0040], 3& 2.542 & -314.55 &  -2.46 & .0550 & 29.38\\
             & [0000], 1& 2.390 & -294.33 & -29.97 & .0450 & 23.32\\
\hline
$Fe_{31}Si$, & [1000], 8& 2.181 & -269.44 & -38.68 & .0497 & 28.27\\
a=21.604 a.u.& [0003], 8& 2.413 & -297.89 & -35.15 & .0471 & 25.55\\
             & [0100], 6& 2.343 & -288.51 & -31.31 & .0446 & 24.46\\
             & [0020], 6& 2.385 & -294.06 & -16.92 & .0468 & 25.32\\
             & [0000], 2& 2.489 & -308.13 & -19.72 & .0485 & 26.44\\
             & [0000], 1& 2.389 & -294.81 & -13.56 & .0451 & 24.28\\
\hline
$Fe_{15}Si$, & [1003], 8& 2.262 & -277.20 & -29.11 & .0594 & 32.38\\
2a=21.585a.u.& [0200], 3& 2.353 & -289.52 & -21.64 & .0456 & 24.44\\
             & [0040], 3& 2.432 & -298.87 &   5.49 & .0509 & 27.31\\
             & [0000], 1& 2.536 & -313.82 & -13.89 & .0496 & 26.83\\
\end{tabular}
\end{table}

\end{document}